\documentclass[reprint,superscriptaddress,prb]{revtex4-2}
\usepackage{amsmath}%
\usepackage{amsfonts}%
\usepackage{amssymb}%
\usepackage{feynmp}%
\usepackage{graphicx}%
\usepackage{setspace}%
\usepackage{comment}%
\usepackage{dsfont}%
\usepackage{color}%
\usepackage[pdftex,colorlinks=true,linkcolor=blue,citecolor=blue]{hyperref}

\usepackage{soul}

\newcommand{\vk}{\Pi}
\newcommand{\vkk}{{\bf \Pi}}

\begin{document}

\title{Higher-order topological pumping and its observation in photonic lattices}
\author{Wladimir A. Benalcazar}
\thanks{these authors contributed equally}
\affiliation{Department of Physics, The Pennsylvania State University, University Park, PA 16801}
\author{Jiho Noh}
\thanks{these authors contributed equally}
\affiliation{Department of Physics, The Pennsylvania State University, University Park, PA 16801}
\author{Mohan Wang}
\thanks{these authors contributed equally}
\affiliation{Department of Electrical and Computer Engineering, University of Pittsburgh, Pittsburgh, PA, USA.}
\author{Sheng Huang}
\thanks{these authors contributed equally}
\affiliation{Department of Electrical and Computer Engineering, University of Pittsburgh, Pittsburgh, Pennsylvania, USA.}
\author{Kevin P. Chen}
\affiliation{Department of Electrical and Computer Engineering, University of Pittsburgh, Pittsburgh, Pennsylvania, USA.}
\author{Mikael C. Rechtsman}
\affiliation{Department of Physics, The Pennsylvania State University, University Park, PA 16801}

\date{\today}

\begin{abstract}
The discovery of the quantization of particle transport in adiabatic pumping cycles of periodic structures by Thouless [Phys. Rev. B {\bf 27}, 6083 (1983)] linked the Chern number, a topological invariant characterizing the quantum Hall effect in two-dimensional electron gases, with the topology of dynamical periodic systems in one dimension. Here, we demonstrate its counterpart for higher-order topology. Specifically, we show that adiabatic cycles in two-dimensional crystals with \emph{vanishing dipole moments} (and therefore zero overall particle transport) can nevertheless be topologically nontrivial. These cycles are associated with higher-order topology and can be diagnosed by their ability to produce corner-to-corner transport in certain metamaterial platforms. We  experimentally verify the corner to corner transport associated with this topological pump by using an array of photonic waveguides adiabatically modulated in their separations and refractive indices. By mapping the dynamical phenomenon demonstrated here from two spatial and one temporal dimensions to three spatial dimensions, our observations are equivalent to the observation of chiral hinge states in a three-dimensional second-order topological insulator.
\end{abstract}

\maketitle

\section{Introduction}
Topological phases of matter exhibit quantized transport properties and robust unconventional states at their boundaries~\cite{HasanKane10,QiZhangreview11}. Initially, they were classified according to whether they obey time-reversal, chiral, or particle-hole symmetries~\cite{altland1997}. More recently, their classification has been refined and enriched by additionally considering crystalline symmetries. In particular, higher-order topological phases in crystalline systems exhibit gapped bulk and boundaries that protect gapless states only at the boundaries of their boundaries, i.e., at their corners or hinges~\cite{benalcazar2014,benalcazar2017quad,benalcazar2017quadPRB,song2017,langbehn2017, schindler2018}. An $n$th-order topological phase in $d$ dimensions protects gapless states in its $(d-n)$-dimensional boundaries.

Higher-order topological phases with protected corner states have been realized in several metamaterial platforms~\cite{noh2018,peterson2018,imhof2018,serragarcia2018,xie2018,khanikaev2018b,xue2018, mittal2019,kempkes2019,bao2019,khanikaev2019,xue2019,xue2019b}. In three-dimensions(3D), the hinge-localized states of second-order topological phases are helical or chiral in nature~\cite{benalcazar2017quadPRB,song2017,schindler2018}. Phases with helical hinge states have been observed in bismuth~\cite{schindler2018a}. Chiral hinge states, however, remain unfound or unrealized, although their existence has been proposed in magnetic axion insulators~\cite{yue2019} and superconductors~\cite{peng2019}.

A second central, though as yet unobserved phenomenon in higher-order topological band theory is the higher-order counterpart of a Thouless pump \cite{thouless1983}.  According to the hierarchical dimensional structure of topological phases, a 2D Chern insulator is connected to a Thouless pump in 1D via dimensional reduction~\cite{thouless1983}. The same dimensional reduction leads to an equivalence between the 4D quantum Hall effect~\cite{zhang2001} and a topological pump that adiabatically connects topological and trivial $\mathbb{Z}_2$ time-reversal invariant insulators in 3D~\cite{qi2008}. More recently, dimensional reduction has also been used to relate the 4D quantum Hall effect to 2D topological pumps over 2D systems~\cite{kraus2013, petrides2020}, features of which have been realized experimentally in photonic~\cite{zilberberg2018} and ultracold atomic~\cite{lohse2018} systems.

The higher-order counterpart of this hierarchical structure establishes an equivalence between 3D second-order topological phases having chiral hinge states and a topological pump that adiabatically connects the trivial atomic limit phase and a second-order obstructed atomic limit (OAL) phase in  2D~\cite{benalcazar2017quadPRB,song2017,petrides2020}.

In this work, we present the first realization of such a higher-order topological pump. We show that this pump manifests corner-localized states that cross the bulk band gap and can be used to adiabatically transport energy from one corner of the structure to its opposite one in a 2D lattice that evolves periodically and adiabatically. We implement this model in an array of photonic waveguides modulated in their separations and refractive indices and use it to demonstrate in an experiment that light is indeed transported as predicted. By dimensionally extending the proposed system from two spatial and one temporal dimensions to three spatial dimensions, our results are the dimensionally-reduced equivalent to the observation of chiral hinge states in 3D second-order topological phases. 

\begin{figure*}[t]
\includegraphics[width=2\columnwidth]{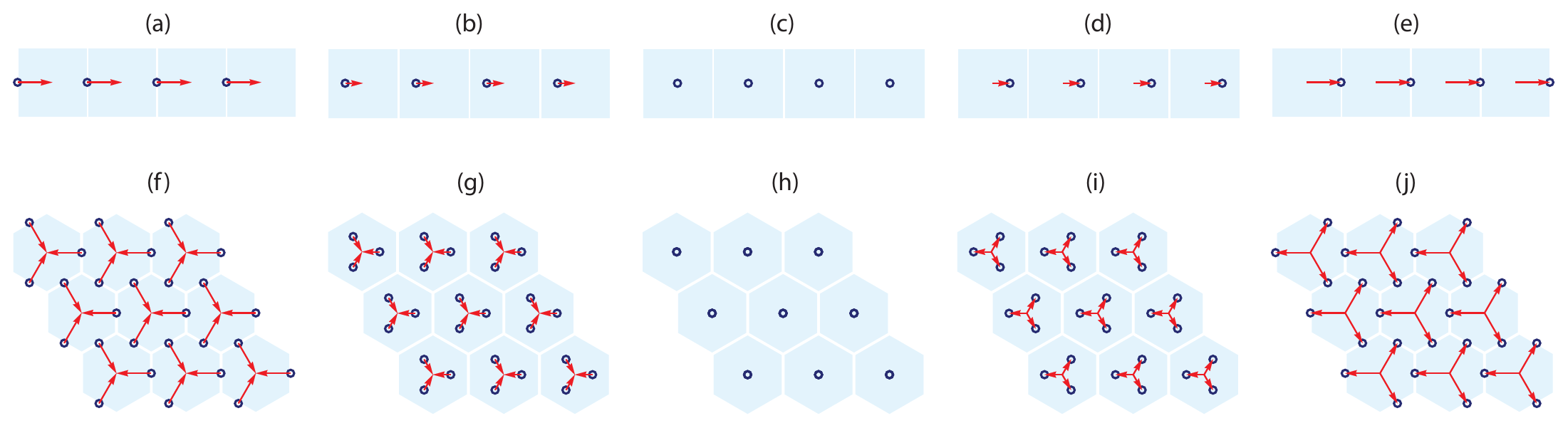}%
\caption{Schematic of the evolution of the Wannier centers (hollow circles) during one cycle of an adiabatic topological pump. [(a)-(e)] A first-order (Thouless) pump. Each unit cell has one Wannier center. [(f)-(j)] The proposed second-order topological pump. Each unit cell has three Wannier centers. In both processes, periodic boundary conditions are adopted. The configurations of Wannier centers at the beginning and end of the cycles are the same. The red arrows indicate the direction of Wannier center flow toward to (first half of cycle) or away from (second half of cycle) their assigned unit cell centers.}
\label{fig:WannierCenters}
\end{figure*}

The rest of the paper is organized as follows: In Sec.~\ref{sec:pumps}, we describe and compare first-order topological pumps and the simplest form of higher-order topological pump, a second-order topological pump, from the perspective of Wannier center flow. In Sec.~\ref{sec:model}, we describe a specific model for the realization of a second-order topological pump, starting from a tight-binding model that reproduces the desired Wannier center flow and showing how it can be implemented in an array of photonic waveguides. In Sec.~\ref{sec:experiment} we first show full wave simulations of the proposed pump that demonstrate that adiabaticity is achieved for the chosen set of parameters for the experiment, and then present the experimental realization of the pump itself. Specifically, we show that light is transported from corner to corner via the bulk of the photonic array, as theoretically expected for a higher-order topological pump. Section~\ref{sec:analysis} presents a detailed analysis of the Brillouin zone (BZ) and the topological characterization and protection of the pump. This section also draws a connection between the proposed pump and a 3D second-order topological phase protected by inversion symmetry and, via this connection, discusses the classification and protection of the topological phase. Section~\ref{sec:discussion} presents a discussion and our conclusions.

\section{First and second-order pumps}
\label{sec:pumps}
In the conventional Thouless pump~\cite{ricemele1982,thouless1983}, an insulator with discrete translation symmetry adiabatically evolves in a periodic fashion leading to a quantization of the electron transport per cycle. The transport can be tracked by following the dipole moment in a crystal as the cycle progresses~\cite{vanderbilt1993}. The change of the dipole moment over a cycle is a topologically protected integer equal to the Chern number of the energy bands calculated in the 2D manifold $(k,\theta)$ spanned by the crystal momentum $k$ over the 1D Brillouin zone and the adiabatic parameter $\theta$ over the cycle~\cite{qi2008}.

Figures~\ref{fig:WannierCenters}(a)-\ref{fig:WannierCenters}(e) illustrate such a process by showing the Wannier centers of a 1D lattice at five stages of a pumping cycle with Chern number $C=1$.
There is a single Wannier center per unit cell. The dipole moment is thus proportional to the position of the Wannier centers relative to the centers of their unit cells~\cite{vanderbilt1993}, indicated in Fig.~\ref{fig:WannierCenters} by arrows. 
The overall effect of the cycle is to transport a Wannier center from left to right by one unit cell, which amounts to the quantization of particle (i.e., Wannier center) transport~\cite{vanderbilt1993}.

The robust quantized transport of Thouless pumps has been observed in lattices of ultra-cold atoms~\cite{lohse2015,nakajima2016}. Thouless pumps in $d$ spatial dimensions manifest states that cross the energy gap and localize on their $(d-1)$-dimensional boundaries. These states have been exploited to transport energy from one edge of a sample to the opposite one in several metamaterial platforms~\cite{kraus2012,verbin2015,grinberg2020}.

Figures~\ref{fig:WannierCenters}(f)-\ref{fig:WannierCenters}(j) illustrate the cycle of the simplest form of higher-order topological pump, a second-order topological pump.  
Each unit cell of the 2D lattice has three Wannier centers. The cycle is $C_3$-symmetric throughout, breaking $C_2$ symmetry at all points except two, at which two topological phases protected by $C_2$ symmetry exist: (i) a 2D second-order topological phase at the beginning and end of the cycle, with Wannier centers in the middle of the edges of the hexagonal unit cells [Figs.~\ref{fig:WannierCenters}(f) and \ref{fig:WannierCenters}(j)], and (ii)  a 2D trivial phase, with three Wannier centers at the center of each unit cell [Fig.~\ref{fig:WannierCenters}(h)]. Note that the overall dipole moment is zero throughout the cycle and, thus, there is no net particle transport per cycle. This is a necessary property of higher-order topological pumps.
Accordingly, these pumps have vanishing Chern numbers in their spatio-temporal manifolds $({\bf k_1},\theta)$ or $({\bf k_2},\theta)$, where ${\bf k_1}$, ${\bf k_2}$ are crystal momenta along inequivalent noncontractible loops in the 2D BZ, and $\theta$ is the adiabatically varying parameter over the cycle. This implies that a system with only edges and no corners will not have protected states that cross the band gap. In the presence of corners, however, protected in-gap states do cross the bulk band gap in the pump of Fig.~\ref{fig:WannierCenters}(f-j). Such corner states can be used to transport energy among them.

\section{Description of the model}
\label{sec:model}
We now describe in detail the minimal tight-binding model that implements the second-order topological pump of Sec.~\ref{sec:pumps} [Fig.~\ref{fig:WannierCenters}(f-j)] and propose its implementation in a 2D photonic waveguide array.

\begin{figure*}[t]
\includegraphics[width=2\columnwidth]{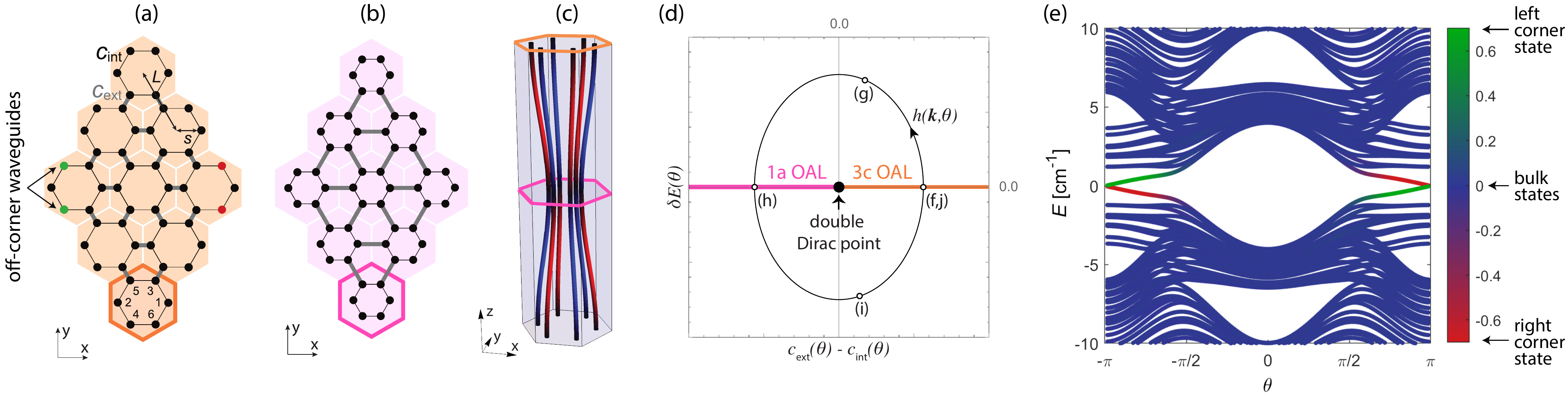}%
\caption{Schematic of the experimental setup for a higher-order topological pump in a crystalline array of photonic waveguides. [(a) and (b)] Cross-sectional profiles  [$(x,y)$ plane cuts] of the array. Dots represent waveguides; gray thick (black thin) lines represent couplings between (within) unit cells with amplitude $c_\text{ext}$ ($c_\text{int}$). The configurations in (a) and (b) are in the $3c$ OAL phase and $1a$ OAL phase (both protected by $C_6$ symmetry), respectively. The orange and pink hollow hexagons indicate a unit cell, and the numbers indicate the basis for the Bloch Hamiltonian in Eq.~\eqref{eq:Hamiltonian}. $L$ is the separation between unit cells, and $s$ is the distance from the center of the unit cell to any of its waveguides. In (a), the green and red ``off-corner" waveguides indicate the two waveguides where the corner topological states have most of their support (for a full plot of the topological corner states, see Appendix~\ref{si:section:topophase}).
(c) A unit cell for the full 3D structure whose separation $s(z)$ and refractive index $\delta n(z)$ (represented by the color gradient in blue and red) are adiabatically modulated as a function of $z=15(\theta+\pi)/2\pi$ cm over a period, $\theta \in [-\pi,\pi)$. (d) Schematic trajectory of the Hamiltonian Eq.~\eqref{eq:Hamiltonian} in parameter space as a function of the adiabatic variable $\theta$. The letters in parentheses indicate the Wannier center configurations as labeled in Fig.~\ref{fig:WannierCenters}. Any closed trajectory that encloses the double Dirac point describes a higher-order topological pump. (e) Tight-binding cross-sectional spectrum as a function of the adiabatic parameter $\theta$ for the structure in (a), (b), and (c). Green, red, and blue colors indicate the support of the states at either the left topological corner state, the right topological corner state, or bulk states, respectively.}
\label{fig:schematic}
\end{figure*}

Specifically, consider the schematic of the three-dimensional photonic structure shown in Fig.~\ref{fig:schematic}.  
At any fixed value of $z$ (the spatial coordinate along the waveguide axis),  the photonic structure is a 2D crystalline array of waveguides in the $(x,y)$ plane with six waveguides per unit cell [Fig.~\ref{fig:schematic}(a), \ref{fig:schematic}(b), and \ref{fig:schematic}(c)]. Each waveguide binds only the lowest-energy transverse electromagnetic (TEM) mode, which has an elliptical (almost circular) profile and evanescently couples to its neighbor modes. The couplings are thus real valued and exponentially decrease with separation between waveguides. We approximate them to be nonvanishing only among nearest neighbors~\footnote{Including additional couplings to waveguides farther away will not preclude topological pumping because the Wannier centers are still pinned to either the center of the unit cell (for the trivial configuration) or at the three middle of the edges of the unit cell (for the topological configuration) as long as $C_6$ symmetry is preserved.}. During the pumping process, we will need to adiabatically modulate the couplings and on-site energies of these modes. To modulate the couplings, we slowly vary the separations of the waveguides in the $(x,y)$ plane as a function of $z$. Similarly, to modulate their on-site energies, we modulate their refractive indices as a function of $z$. These two modulations are schematically represented in Fig.~\ref{fig:schematic}(c) for one unit cell. Here, we exploit the fact that the spatial evolution of the optical beam parallel to the waveguides' axis directly maps onto the temporal evolution described by the Schr\"odinger equation of a quantum particle, via the paraxial approximation (Appendix~\ref{si:section:paraxial}). Thus, this slow variation of waveguide parameters as a function of $z$ is equivalent to an adiabatic temporal evolution of the system in the $(x,y)$ plane. A tight-binding Bloch Hamiltonian for this process is given by
\begin{align}
h({\bf k},\theta) &= c_\text{ext}(\theta) h_\text{ext}({\bf k}) + c_\text{int}(\theta) h_\text{int} + \delta E(\theta) \Pi,
\label{eq:Hamiltonian}
\end{align}
where $\theta$ is the adiabatic parameter (which varies linearly with $z$), ${\bf k}=(k_x,k_y)$ is the crystal momentum in the $(x,y)$ plane, $c_\text{ext}$ ($c_\text{int}$) represents the coupling amplitudes between nearest-neighbor waveguides between (within) unit cells (which exponentially decrease with separation), and $\delta E$ are the on-site energies (which vary with refractive index). The terms $h_\text{ext}({\bf k})$, $h_\text{int}$, and $\Pi$ are $6 \times 6$ matrices with entries corresponding to the six waveguides numbered in Fig.~\ref{fig:schematic}(a), and take the form $h_\text{ext}({\bf k})=\oplus_{i=1}^3[\cos({\bf k \cdot a_i})\sigma_x+\sin({\bf k \cdot a_i})\sigma_y]$, with ${\bf a_1}=(1,0)$, ${\bf a_{2,3}}=(\pm 1/2,\sqrt{3}/2)$ primitive lattice vectors in the $(x,y)$ plane, $h_\text{int}$ has entries $[h_\text{int}]_{mn}=1$ for nearest-neighbor waveguides $m$ and $n$ within the same unit cell and 0 otherwise, and $\Pi = \sigma_z \oplus (-\sigma_z) \oplus \sigma_z$. Here $\sigma_{i=x,y,z}$ are the Pauli matrices.

In the absence of modulation of the refractive index we have $\delta E(\theta)=0$, and the Hamiltonian Eq.~\eqref{eq:Hamiltonian} is, for all values of $\theta \in [0,2\pi)$, (i) $C_6$ symmetric, obeying $\hat{r}_6 h({\bf k}, \theta) \hat{r}^\dagger_6=h(R_6 {\bf k},\theta)$, where $\hat{r}_6$ is the $C_6$ rotation operator acting on the degrees of freedom within a unit cell [i.e., the six numbered waveguides in Fig.~\ref{fig:schematic}(a)] and $R_6$ rotates the crystal momentum by $60^\circ$, and (ii) chiral symmetric, $\Pi h({\bf k},\theta) \Pi = -h({\bf k},\theta)$. Depending on the ratio of the couplings $c_\text{ext}/c_\text{int}$, the structure is in one of two phases protected by $C_2$ symmetry; when $c_\text{ext}/c_\text{int}>1$, the structure is in the OAL phase with Wannier centers as in Figs.~\ref{fig:WannierCenters}(f) and \ref{fig:WannierCenters}(j) which is characterized by the presence of midgap corner-localized zero-energy states~\cite{noh2018}. We refer to this phase as the $3c$ OAL phase. On the other hand, when $c_\text{ext}/c_\text{int}<1$, the structure is in the trivial atomic limit phase with Wannier centers as in  Fig.~\ref{fig:WannierCenters}(h) and hosts no boundary states (Appendix~\ref{si:section:topophase}). We refer to this phase as the $1a$ OAL phase. These two phases are separated by a critical point at $c_\text{ext}/c_\text{int}=1$, where two of the bulk bands of the 2D crystal below zero energy close their gap with two bands above zero energy at the ${\bf \Gamma}$ point of the BZ. For adiabatic pumping, we need to avoid the phase transition as we modulate the separations between waveguides. For that purpose, we additionally modulate the refractive index. This modulation amounts to the third term in the Hamiltonian Eq.~\eqref{eq:Hamiltonian} proportional to $\delta E(\theta)$, which breaks $C_6$ symmetry down to only $C_3$ symmetry and allows for an adiabatic flow of the Wannier centers. Since the added term is proportional to the chiral operator, $\Pi$, which anticommutes with the rest of the Hamiltonian, a gap will be guaranteed if $\delta E(\theta) \neq 0$ when $c_\text{ext}/c_\text{int}=1$. A schematic trajectory of the adiabatic pump in Eq.~\eqref{eq:Hamiltonian} in parameter space is shown in Fig.~\ref{fig:schematic}(d). In that figure, a nontrivial topological pump is protected for any closed trajectory of $h({\bf k},\theta)$ that encloses the double Dirac point at $(\delta E, c_\text{ext}-c_\text{int})=(0,0)$.
For concreteness, we choose the following modulation of couplings and refractive indices,
\begin{align}
c_\text{ext}(\theta) &= C e^{-\kappa [L-2 s(\theta)]},\quad
c_\text{int}(\theta) = C e^{-\kappa s(\theta)},\nonumber\\
\delta E (\theta) &= \epsilon \delta n_0 \sin(\theta),
\label{eq:modulation}
\end{align}
where $s(\theta) =L/3-A\cos(\theta)$ (for $A<L/3$) is the separation between neighboring waveguides within a unit cell, $L=50$ $\mu$m is the separation between unit cells [Fig.~\ref{fig:schematic}(a)], $A=2.2$ $\mu$m and $\delta n_0=0.5\times 10^{-4}$ are the amplitudes of modulation of the separation $s$ and the refractive index, respectively, and $\kappa=0.19$~$\mu$m$^{-1}$, $C=77$~cm$^{-1}$, and $\epsilon = 1.469 \times 10^4$~cm$^{-1}$ are experimental parameters at $\lambda=1555$~nm.

A plot of the energy bands for this process is shown in Fig.~\ref{fig:schematic}(e). Although bulk and edges are gapped (blue lines), there is a pair of gapless states that cross the energy band, which localize at $120^\circ$ corners (red and green lines) [most of the support of the corner states is in the ``off-corner" waveguides marked by green and red colors in Fig.~\ref{fig:schematic}(a)], see Ref.~\cite{noh2018} and Appendix~\ref{si:section:topophase}.
These states do not hybridize at the beginning or end of the cycle because they localize at opposite corners.

\section{Experimental implementation and observation}
\label{sec:experiment}
\begin{figure*}[t]
\includegraphics[width=2\columnwidth]{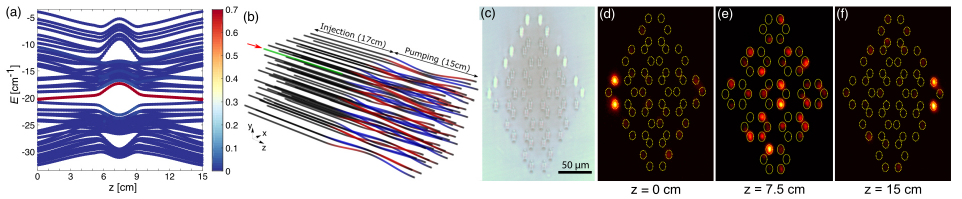}
\caption{Simulation (a) and experimental observation [(b)-(f)] of the second-order topological pump with Bloch Hamiltonian Eq.~\eqref{eq:Hamiltonian} with the setup of Fig.~\ref{fig:schematic}. (a) Spectrum during one pumping cycle using the full wave propagation simulation. The color map indicates the amplitude of the projection of the propagating beam onto each of the instantaneous eigenstates of the system. (b) Schematic of the waveguide array of the second-order topological pump (dimensions not to scale). The experiment consists of two successive stages along the propagation direction, $z$. The green waveguide indicates the auxiliary waveguide placed in close proximity to one of the 120$^{\circ}$  corners in the 17cm-long injection segment. The refractive indices of the rest of the waveguides are represented by the color gradient in blue and red. (c) Input facet of the pumping segment. The radii of the major and minor axes of the waveguides are 5.35 and 3.5 $\mu$m, the separation between unit cells is $L=50$ $\mu$m. The input facet is in the $3c$ OAL phase, shown schematically in Fig.~\ref{fig:schematic}(a). (d) Beam profile at the output facet of the injection segment, where the topological corner mode is prepared. (e) Beam profile at the center of the pumping segment. (f) Beam profile at the output. In (d)-(f), yellow lines are overlapped to indicate the positions of the waveguides at each facet.}
\label{fig:experiment}
\end{figure*}

From the spectrum of Fig.~\ref{fig:schematic}(e) it follows that an initial beam that occupies one of the corner eigenstates in the waveguide array at $\theta=-\pi$ will delocalize into the bulk as it adiabatically approaches $\theta=0$ but will emerge at the opposite corner at $\theta=\pi$. This will occur if the beam does not occupy bulk states other than the one adiabatically connected with the topological corner states. In the past, it has been shown that such an adiabatic transport can be achieved in optical waveguide arrays~\cite{kraus2012}. Here, we use this method to probe the corner to corner transport, which amounts to verifying the bulk-boundary correspondence of this second-order topological pump. Fig.~\ref{fig:experiment}(a) shows the instantaneous eigenvalues at each value of $z$ for a simulation of beam propagation in a system with a modulation adiabatically deformable to that of Eq.~\eqref{eq:modulation}. Unlike the tight-binding simulation leading to Fig.~\ref{fig:schematic}(e), the simulation for Fig.~\ref{fig:experiment}(a) is carried out using the beam propagation method (BPM), which simulates the evolution of the field $\psi({\bf r},z)$ within the paraxial approximation (with no tight-binding approximation assumed). The parameter values and results of this simulation are detailed in Appendix~\ref{si:section:sim}.

The color in Fig.~\ref{fig:experiment}(a) indicates the amplitude of the projection of a propagating beam into the instantaneous eigenstates at each value of $z$. Adiabaticity is evidenced by the fact that the propagating beam does not occupy other eigenstates close to the beam's frequency at any point of the pump cycle. We observe that the beam, injected initially at the left corner, occupies the bulk as the cycle progresses and finally appears at the right corner at the end of the cycle. The power transmitted to the right corner state was $99\%$ of the original incident power on the left corner state for a simulated sample of 15 cm.

We fabricated a sample that reproduced the modulation scheme of the BPM simulations to observe the corner-to-corner transport experimentally. We do this by fabricating two waveguide arrays, an ``injection segment" that populates the corner topological state, and a ``pumping segment", in which the waveguides are modulated to produce the pump [Fig.~\ref{fig:experiment}(b)]. Figures~\ref{fig:experiment}(c)-\ref{fig:experiment}(f) show microscope images of the sample's input facet of the pumping segment, as well as of the optical wave functions at the beginning, middle, and end of the pumping cycle, respectively. 

At the start of the pumping cycle, the beam occupies the topological corner state. To generate that corner state, we use the injection segment of the sample. There, we indirectly excite the topological corner mode by using an auxiliary waveguide close to the left corner of a constant (i.e., with no modulation) waveguide array set to the $3c$ OAL phase. For this injection segment, of length 17 cm, we set a refractive index of $\Delta n=3.00 \times 10^{-3}$ for both the waveguide array and the auxiliary waveguide, and a ratio of  $L/s=2.64$. The topological corner mode, being degenerate with the mode in the auxiliary waveguide, weakly couples to it. We input the beam into the auxiliary waveguide and let it leak into the topological corner mode of the array during the entire 17 cm of this injection segment. Figure~\ref{fig:experiment}(d) shows the diffracted light measured at the output facet of the injection segment. From this output facet, only the two off-corner waveguides that comprise the vast majority of the left corner topological state are directly butt-coupled to the off-corner waveguides at the left corner of the input facet of the pumping segment. 

In the pumping segment of the sample, the positions and the refractive indices of the waveguides were modulated according to the parameters of the full-wave simulation, see Appendix~\ref{si:section:sim}. The $L/s$ ratio at the input and output facets was set to 2.64, while halfway through propagation ($z=7.5$ cm) it was set to 3.32. $\delta n_0$ was set to $0.5\times 10^{-4}$.  After performing the experiment in the full sample, we cleaved it at $z=7.5$ cm, allowing us to verify that the state occupies the bulk of the array halfway through the pump cycle [Fig.~\ref{fig:experiment}(e)], while the output facet of the pumping cycle shows that the beam was transported to the opposite topological corner mode [Fig.~\ref{fig:experiment}(f)], as expected.

\section{Brillouin zone analysis and topological protection}
\label{sec:analysis}
Having demonstrated experimentally the signatures associated with the proposed second-order topological pump in bosonic systems, we now describe in detail the topological origin and symmetry protection in the model. In doing so, we also draw connections with its equivalent 3D higher-order topological phase and its protected hinge-localized chiral modes.

\subsection{Topological classifications, band representations, and Wannier flow}
Crystalline symmetries can fix the positions of Wannier centers to maximal Wyckoff positions of the unit cell, generating OAL phases~\cite{bradlyn2017}. As stated in Sec.~\ref{sec:model}, the proposed second-order topological pump, Eq.~\eqref{eq:Hamiltonian}, adiabatically connects two 2D OAL phases protected by $C_6$ symmetry by means of breaking $C_6$ symmetry down to only $C_3$ symmetry at all intermediate points between those two OAL phases. This connection is only possible if $C_2$ symmetry is crucial in protecting the two OAL phases. In this section we show, by using the theory of band representations~\cite{bradlyn2017}, that $C_2$ symmetry indeed protects the 2D OAL phases and that the Hamiltonian Eq.~\eqref{eq:Hamiltonian} has the symmetry representations compatible with the Wannier flow of Figs.~\ref{fig:WannierCenters}(f)-\ref{fig:WannierCenters}(j). 

Let us start by describing the symmetry indicator invariants that will diagnose the 2D OAL phases. We denote the eigenvalues of the rotation operator $\hat{r}_n$ as
\begin{align}
\vk^{(n)}_p = e^{2\pi i (p-1)/n}, \quad \text{for } p=1,2,\ldots n.
\end{align}
At high-symmetry points (HSPs) ${\vkk}$ of the BZ that are invariant under $C_n$ rotation symmetry we define the integer invariants~\cite{teo2013,benalcazar2014,benalcazar2019fillinganomaly,ortix2018}
\begin{align}
[\vk^{(n)}_p] \equiv \# \vk^{(n)}_p-\#\Gamma^{(n)}_p \quad \in \mathbb{Z},
\label{eq:RotationInvariants}
\end{align}
where $\# \vk^{(n)}_p$ is the number of states below zero energy with eigenvalue $\vk^{(n)}_p$. In $C_6$ symmetric and $C_3$ symmetric crystals, the classifications of topological phases are indicated by the following indices~\cite{benalcazar2019fillinganomaly}
\begin{align}
\chi_\mathcal{T}^{(6)}&=( [M^{(2)}_1],[K^{(3)}_1] ),\nonumber\\
\chi_\mathcal{T}^{(3)}&=( [K^{(3)}_1], [K^{(3)}_2] ),
\label{eq:ClassificationIndicesTRS}
\end{align}
\begin{table}[t]
	\centering
	\begin{tabular}{|c|c|c|c|}
		\hline 
		OAL phase & $C_2$ evals ${\bf \Gamma}$ & $C_2$ evals ${\bf M}$ & $\chi_\mathcal{T}^{(6)}$\\
		\hline\hline
		$1a$ & $\#\Gamma_1=2$ & $\#M_1=2$ & $(0,0)$\\
		& $\#\Gamma_2=1$ & $\#M_2=1$ & \\
		\hline
		$3c$ & $\#\Gamma_1=0$ & $\#M_1=2$ & $(2,0)$\\
		& $\#\Gamma_2=3$ & $\#M_2=1$ & \\
		 \hline
	\end{tabular} 
	\caption{$C_2$ symmetry representations and invariants of the lowest three energy bands for the $1a$ OAL and $3c$ OAL phases ($\theta=0$ and $\pi$ in Eq.\ref{eq:Hamiltonian}, respectively).}
	\label{tab:C6_inducedC3InvariantsFrom3c}
\end{table}
\begin{figure*}[t]
\includegraphics[width=2\columnwidth]{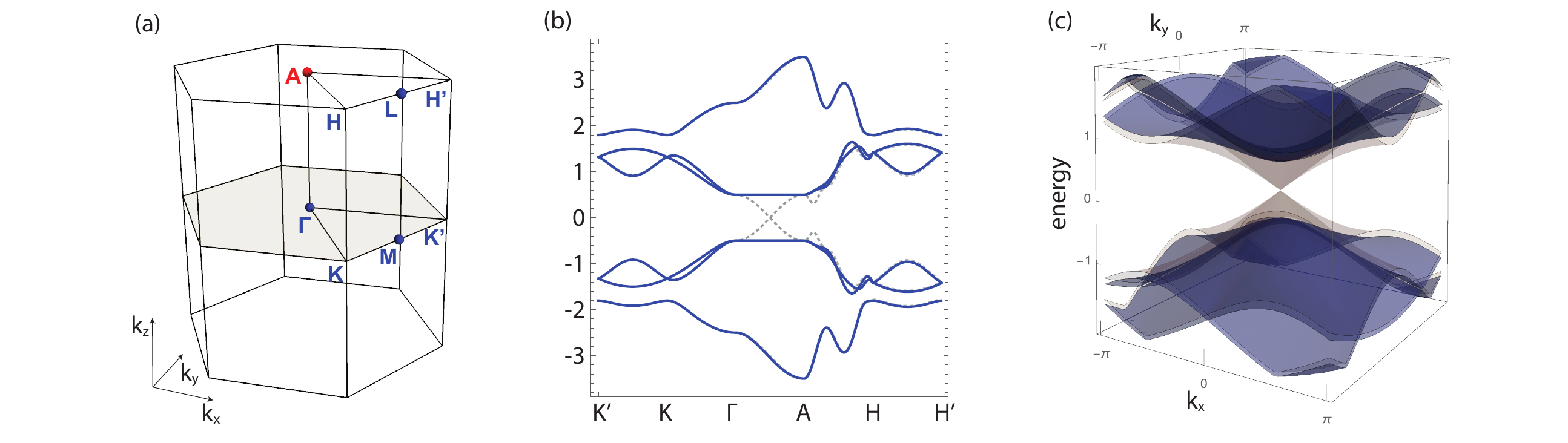}%
\caption{Equivalence between the second-order topological pump, Eq.~\eqref{eq:Hamiltonian}, and a 3D higher-order topological system with chiral hinge states upon the replacement $\theta \rightarrow k_z$. (a) Brillouin zone and its high-symmetry points. ${\bf \Gamma}$, ${\bf M}$, ${\bf A}$, and ${\bf L}$ are invariant points under inversion.  There is a double band inversion at ${\bf A}$ (red dot). (b) The energy bands along specific high-symmetry lines. (c) Energy bands as a function of $(k_x,k_y)$ for a fixed value of $k_z=\pi/2$. In (b) and (c), gray and blue bands correspond to $\delta E=0$ and $\delta E>0$, respectively.}
\label{fig:HOTI}
\end{figure*}
respectively. In Eq.~\eqref{eq:ClassificationIndicesTRS}, $[M^{(2)}_1]$ is an invariant protected by $C_2$ symmetry while $[K^{(3)}_1]$ and $[K^{(3)}_2]$ are protected by $C_3$ symmetry. We find for the proposed pump Eq.~\eqref{eq:Hamiltonian},
\begin{align}
\chi^{(6)}_\mathcal{T} &= \left\{ \begin{array}{c}
(0,0)\quad \text{for } \theta = 0\\
(2,0)\quad \text{for } \theta = \pi
\end{array}\right.\nonumber\\
\chi^{(3)}_\mathcal{T} &= (0,0) \quad \text{for } -\pi<\theta \leq \pi.
\end{align}
With these classifications in mind, we now show that the two $C_6$-symmetric OAL phases at $\theta = \pi$ and $\theta = 0$ have the Wannier configurations of Figs.~\ref{fig:WannierCenters}(f) and \ref{fig:WannierCenters}(h), respectively. We do this by solving the inverse problem, i.e., by inducing the representations at the $C_2$ invariant and $C_3$ invariant points of the BZ from the Wannier center configurations~\cite{bradlyn2017}. In the $1a$ OAL phase, the site symmetry representation is the same at all rotation invariant points, leading to trivial invariants $\chi_\mathcal{T}^{(6)}=(0,0)$ (Table~\ref{tab:C6_inducedC3InvariantsFrom3c}), which is the configuration of the pump at $\theta=0$. In contrast, in the $3c$ OAL phase, the band representations of $C_3$ and $C_2$ are given by~\cite{cano2018}
\begin{align}
\rho_G^{\bf k}(C_3)&=\left(\begin{array}{ccc}
0 & 0 & 1\\
1 & 0 & 0\\
0 & 1 & 0
\end{array}\right), \nonumber \\
\rho_G^{\bf k}(C_2)&=\left(\begin{array}{ccc}
e^{i {\bf k.{a}_2}}& 0\\
0 & e^{-i {\bf k.{a}_1}} &0\\
0 & 0 & e^{-i {\bf k.{a}_3}}
\end{array}\right) \rho(C_2).
\label{eq:BandReps}
\end{align}
Since the representations for $C_3$ symmetry are ${\bf k}$ independent [first equation in Eq.~\eqref{eq:BandReps}], the band representations are constant across the $C_3$ invariant points ${\bf \Gamma}$ and ${\bf K}$. Therefore, the $C_3$ symmetry indicator topological invariant is trivial, i.e., $[K_1^{(3)}]=0$.  From the second equation in Eq.~\eqref{eq:BandReps}, the band representations at the two $C_2$ invariant points are
\begin{align}
\rho_G^{\bf \Gamma}(C_2)=I_{3 \times 3}\rho(C_2),\;\;
\rho_G^{\bf M}(C_2)=\begin{pmatrix}
-1 & 0 & 0\\
0 & -1 & 0\\
0 & 0 & 1
\end{pmatrix}\rho(C_2), \nonumber
\end{align}
which leads to the eigenvalues and invariants in Table~\ref{tab:C6_inducedC3InvariantsFrom3c} if the site-symmetry representation is $\rho(C_2)=-1$.

Note that the two OAL phases differ from one another in their $C_2$ symmetry indicator invariant $[M_1^{(2)}]$. If $C_2$ symmetry is broken, the distinction between the phases is lost, and they can be connected adiabatically (i.e., without going through a gap-closing phase transition). The Hamiltonian Eq.~\eqref{eq:Hamiltonian} achieves that connection by a process that breaks $C_6$ symmetry down to $C_3$ symmetry at all points between the two 2D OAL phases (but not at the OAL phases themselves). To claim that such Wannier center connection is possible, we must also ensure that there are no constraints from $C_3$ symmetry that preclude Wannier centers to move freely (since such symmetry is preserved throughout the pump). Indeed, from Eq.~\eqref{eq:BandReps} it follows that the eigenvalues of $C_3$ symmetry at all HSPs are always $\{K_1, K_2, K_3\}$. Being the same across all HSPs, they lead to trivial $C_3$ symmetry indicator invariants, $\chi_\mathcal{T}^{(3)}=(0,0)$. Moreover, these eigenvalues are those of the permutation matrix, which is the form the $C_3$ rotation operator must take for Wannier centers to be free to move in a $C_3$ symmetric fashion (in real space, the $C_3$ rotation operator will permute the Wannier centers upon a 120$^\circ$ rotation). Since the entire process does not close the energy gap at zero energy, the $C_3$ eigenvalues will persist throughout the entire pumping process, allowing for the smooth connection of the $1a$ OAL and $3c$ OAL phases.

\subsection{Equivalence with a 3D HOTI}
\label{sec:Equivalence}
If the adiabatic parameter $\theta$ is interpreted as a crystal momentum in a third direction, $k_z$, the spectrum in Fig.~\ref{fig:schematic}(d) amounts to that of a 3D second-order topological insulator with open boundaries in the $(x,y)$ plane and periodic boundaries along $z$. The in-gap protected states then correspond to hinge-localized chiral states with opposite propagation directions, with the left and right hinges having positive and negative group velocities along $z$, respectively. 

The equivalent 3D crystal must now possess inversion symmetry, $\mathcal{I} h({\bf k}) \mathcal{I}^{-1}=h(-{\bf k})$, for ${\bf k}=(k_x,k_y,k_z)$, to protect the higher-order topological phase~\cite{ortix2018,khalaf2018,khalaf2018b,kooi2018}. This can be seen by the fact that, at the $k_z=0$ and $k_z=\pi$ planes of the 3D BZ, inversion symmetry coincides with the 2D $C_2$ symmetry. At the inversion invariant point ${\bf A}$, the inversion eigenvalues of the energy bands below zero energy are $\{-1,-1,-1\}$, while at all the other three inversion invariant points they are $\{+1,+1,-1\}$ [Fig.~\ref{fig:HOTI}(a)]. Since two eigenvalues change at ${\bf A}$ in comparison to all other HSPs, there is a double band inversion at ${\bf A}$, characteristic of higher-order topological phases. In contrast, having a single band inversion would signal a strong topological insulator phase~\cite{fu2007}.

It is illustrative to see how the modulation of the refractive indices [third term in Eq.~\eqref{eq:Hamiltonian}] amounts to the mass term that causes such double band inversion in the equivalent 3D crystal.  Figures~\ref{fig:HOTI}(b) and \ref{fig:HOTI}(c) show the energy bands for a value of $\delta E=0$ and $\delta E>0$ in gray and blue, respectively, around the momenta ${\bf k^*}=(0,0,\pi/2)$ (we have set the unit cell length along $z$ to 1). Both plots jointly show that, when $\delta E=0$, there is a double Dirac point at ${\bf k^*}$, and that setting $\delta E>0$ gaps them out. From the point of view of the Wannier flow in the (2D+1)-dimensional pump, a term $\delta E>0$ avoids the phase transition between the $1a$ OAL phase and the $3c$ OAL phase by breaking $C_6$ symmetry down to only $C_3$ symmetry, thus enabling a smooth flow of the Wannier centers. The protection of the 3D second-order TI by inversion symmetry maps to a protection due to $C_2T$, where $T$ is the time-reversal operator obeying $T^2=1$.

\subsection{Robustness of the topological states against disorder}
\begin{figure}[t]
\includegraphics[width=\columnwidth]{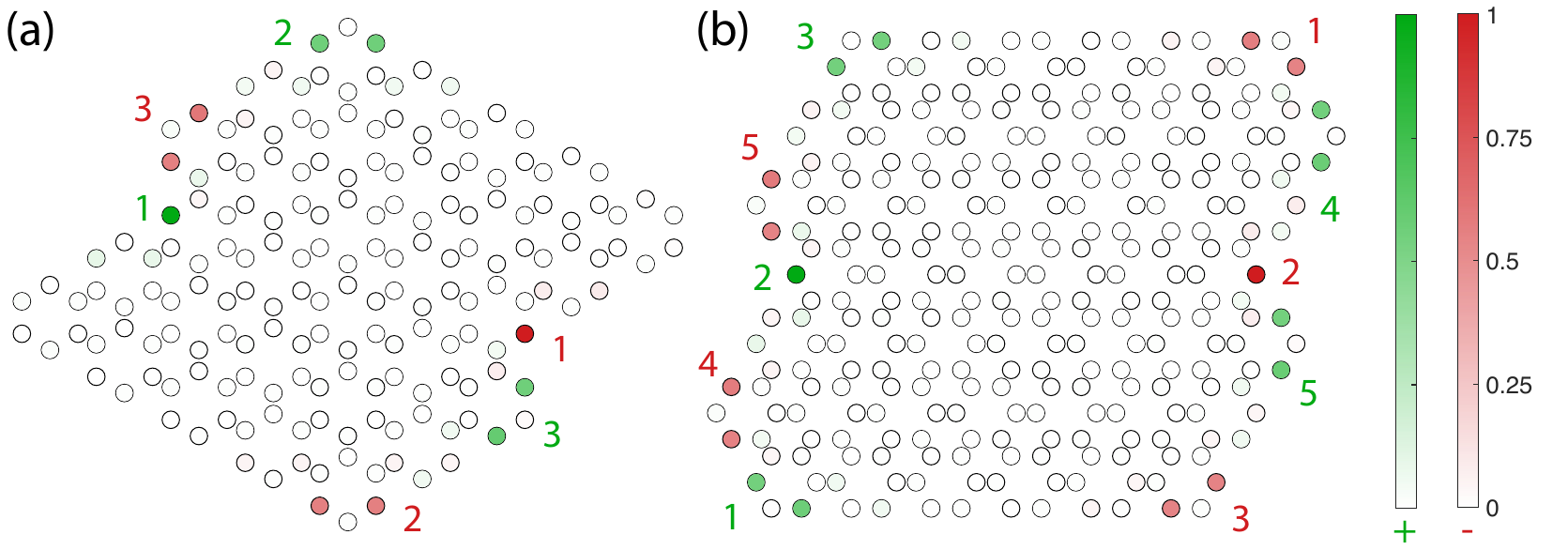}%
\caption{Top view of hinge states in two 3D inversion symmetric lattices in the higher-order topological phase (tight-binding simulations). In both (a) and (b) red and green colors distinguish states with opposite group velocity and the intensity of the color is proportional to the local density of states. The numbers enumerate the pairs of hinge states related by inversion-symmetry. In any nontrivial inversion-symmetric phase, there is an odd number of pairs of hinge states with opposite group velocity, so that at least one pair cannot be gapped out by inversion-symmetric boundary deformations.}
\label{fig:boundaries}
\end{figure}

The second-order topological pump, and hence its equivalent 3D higher-order phase, are characterized by states that traverse the gap. These topological states can be removed by boundary deformations that bring them together in pairs of states with opposite group velocity. As such a pair of states is brought together, it will naturally hybridize, destroying the pair of topological states. However, for any shape of the lattice that preserves inversion symmetry, the nontrivial phase will always present an odd number of pairs of topological states with opposite group velocity and localized at opposite ends of the lattice. Hence, any boundary distortion that preserves inversion symmetry necessarily leaves at least one pair that cannot hybridize, distinguishing from a trivial phase with the same configuration, which will show no topological states (Fig.~\ref{fig:boundaries}). Thus, the classification of these phases is $\mathbb{Z}_2$, and distinguishes topological phases with odd (nontrivial phase) or even (trivial phase) pairs of topological hinge states. The latter can correspond to a case in which two nontrivial phases are connected; each contributes an odd number of pairs, but the system as a whole will have an even pair, all of which can hybridize while preserving the protecting symmetries. 

\section{Discussion and conclusion}
\label{sec:discussion}
We have presented a second-order topological pump in theory and experiment. The pump is equivalent to an intrinsic higher-order topological insulator with hinge states protected by a bulk topological invariant associated with a twofold bulk band inversion. This phase is characterized by an odd number of topological pairs of gapless states related by inversion symmetry, so that boundary deformations that preserve inversion symmetry and attempt to gap out all topological states cannot possibly gap out the last pair. Consequently, unlike first order pumps, which are $\mathbb{Z}$ classified, the present pump is $\mathbb{Z}_2$ classified.

We first stress the fact that the proposed model is not an extrinsic topological phase, i.e., one with a trivial bulk to which nontrivial phases of lower dimension are attached to its boundaries. This is evident from its topological protection, which is characterized by a bulk symmetry-indicator topological invariant -- not a boundary one -- and it is evidenced experimentally in the fact that the pump cycle transported the optical power of a beam from one corner to its opposite one across its bulk, and not across its edges.

It is also worth noting that higher-order pumps should not cause edge-to-edge transport via in-gap states (i.e., their edges should remain gapped). In the proposed system, this is the case as the dipole moment identically vanishes at each instant of the pump.  If edge-to-edge transport were to occur, the dimensionally extended 3D second-order phase would have gapless states on its 2D surfaces into which the hinge states could scatter in the presence of disorder. This distinguishes the present result from that of the 2D pump (dimensionally reduced from 4D) of Ref.~\cite{zilberberg2018}; while the corner states observed in that work correspond to localized states on the hinges of the three-dimensional cube, they are not protected against scattering into the degenerate surface states.

Higher-order topological phases with chiral hinge states remain to be found in condensed matter systems. In photonics, generating chiral hinge states is particularly difficult due to the necessity to break time-reversal symmetry in a 3D bulk. We have circumvented this difficulty via dimensional reduction, by which chiral hinge states in a 3D second-order topological insulator map to the topological pump on a 2D second-order TI, like the one we probe in the present experiment. Thus our experiment effectively provides experimental access to the anomalous chiral hinge states in 3D second-order TIs.

\begin{acknowledgments}
We acknowledge discussions with Thomas Christensen, Tianhe Li, Marius J\"urgensen, Oded Zilberberg, and Ioannis Petrides. W.A.B. thanks the support of the Eberly Postdoctoral Fellowship at the Pennsylvania State University. J.N. acknowledges support by Corning Incorporated Office of STEM Graduate Research.  M.C.R. and J.N. acknowledge the support of the US Office of Naval Research (ONR) Multidisciplinary University Research Initiative (MURI) Grant No. N00014-20-1-2325 on Robust Photonic Materials with High-Order Topological Protection. M.C.R. acknowledges the Packard Foundation under Fellowship No. 2017-66821.  
\end{acknowledgments}

\appendix

\section{Light propagation in the waveguide array and its tight-binding approximation}
\label{si:section:paraxial}
The diffraction of light through the waveguide array of Fig.~\ref{fig:schematic} is governed by the paraxial wave equation
\begin{align}
i\partial_{z}\psi(\textbf{r},z) = \left[-\frac{1}{2k_{0}}\nabla^{2}_{\textbf{r}}-\frac{k_{0}\Delta n(\textbf{r},z)}{n_{0}}\right]\psi(\textbf{r},z),
\label{si:eq:propagation}
\end{align}
where $\psi(\textbf{r},z)$ is the envelope function of the electric field $\textbf{E}(\textbf{r},z)=\psi(\textbf{r},z)e^{i(k_{0}z-\omega t)}\hat{x}$, $k_{0}=2\pi n_{0}/\lambda$ is the wave number within the medium, $\lambda$ is the wavelength of light, $\nabla^{2}_{\textbf{r}}$ is the Laplacian in the transverse $(x,y)$ plane, $\omega=2\pi c/\lambda$, and $\Delta n$ is the refractive index relative to $n_{0}$.
Assuming that only the lowest TEM mode is bound to each waveguide, we may employ the tight-binding approximation
\begin{align}
i\partial_{z}\psi_{i}(z)=-\sum_{j}c_{ij}(\lambda)\psi_{j}(z),
\label{si:eq:coupledmodeEq}
\end{align}
where $\psi_{n}$ is the amplitude of the optical field in the $n$th waveguide, and $c_{ij}(\lambda)$ is the coupling constant between waveguides $i$ and $j$ at wavelength $\lambda$.
For a tight-binding description in the adiabatic regime, we assume the propagation direction $z$ to be constant. Thus, we can explicitly write the $z$ dependence of the propagating modes in Eq.~\eqref{si:eq:coupledmodeEq} as $\psi_n(z)=\psi_n e^{i E z}$. This leads to
\begin{equation}
E \psi_{i}=\sum_{j}c_{ij}(\lambda)\psi_{j},
\label{si:eq:eigensystemEq}
\end{equation}
where $E$ plays the role of energy in the analogous Schr{\"o}dinger equation $H \psi = E \psi$, where $H_{ij}\equiv c_{ij}$.

\section{Higher-order topology of the $C_6$-symmetric OAL phases}
\label{si:section:topophase}
The second-order topological pump described by Eq.~\eqref{eq:Hamiltonian} and Eq.~\eqref{eq:modulation} adiabatically connects two 2D OAL phases protected by $C_2$ symmetry. A detailed account of these two phases can be found in Ref.~\cite{noh2018}. For the sake of completeness, however, here we summarize the essential aspects of these phases. 

These two phases are characterized by the Hamiltonian in Eq.~\eqref{eq:Hamiltonian} with the parameters in Eq.~\eqref{eq:modulation} at values of $\theta=0$ and $\pi$ for the trivial and topological phases, respectively. At these two values of the adiabatic parameter this model has $C_6$ symmetry,
\begin{align}
\hat{r}_6 h({\bf k}) \hat{r}_6^\dagger = h (R_6 {\bf k}),
\end{align}
where
\begin{align}
\hat{r}_6 = \left(\begin{array}{cccccc}
0&\sigma_0&0\\
0&0&\sigma_0\\
\sigma_x&0&0
\end{array}\right),\;\;
R_6 = \left(\begin{array}{cc}
\cos(2\pi/6) & \sin(2\pi/6)\\
-\sin(2\pi/6) & \cos(2\pi/6)
\end{array}\right),
\label{eq:rotation_symmetry}
\end{align}
are the rotation operator acting on the internal degrees of freedom of the unit cell, which obeys $\hat{r}_6^6=1$, and the matrix that rotates the crystal momentum by $2\pi/6$ radians, respectively. The $\sigma_x, \sigma_y, \sigma_z$ are Pauli matrices and $\sigma_0$ is the $2\times 2$ identity matrix. 

As long as $C_6$ symmetry is preserved, the crystalline structure can transition from a topological phase to the trivial phase as we vary the ratio $c_{\text{int}}/c_{\text{ext}}$. 
The corresponding topological phases are
\begin{align}
\chi^{(6)}_\mathcal{T} = \left\{ \begin{array}{c}
(0,0)\quad \text{for } c_{\text{int}}/c_{\text{ext}} > 1\\
(2,0)\quad \text{for } c_{\text{int}}/c_{\text{ext}} < 1
\end{array}\right.
\end{align}
The transition at $c_{\text{int}}/c_{\text{ext}}=1$ occurs by closing the bulk energy gap at the $\bf \Gamma$ point of the BZ. This transition point corresponds to the usual honeycomb lattice, which is well known in the context of graphene to have two Dirac cones. The difference in our formulation resides exclusively in our unit cell definition having six instead of two degrees of freedom (see Fig.~\ref{fig:schematic}).
\begin{figure}[t]
\includegraphics[width=\columnwidth]{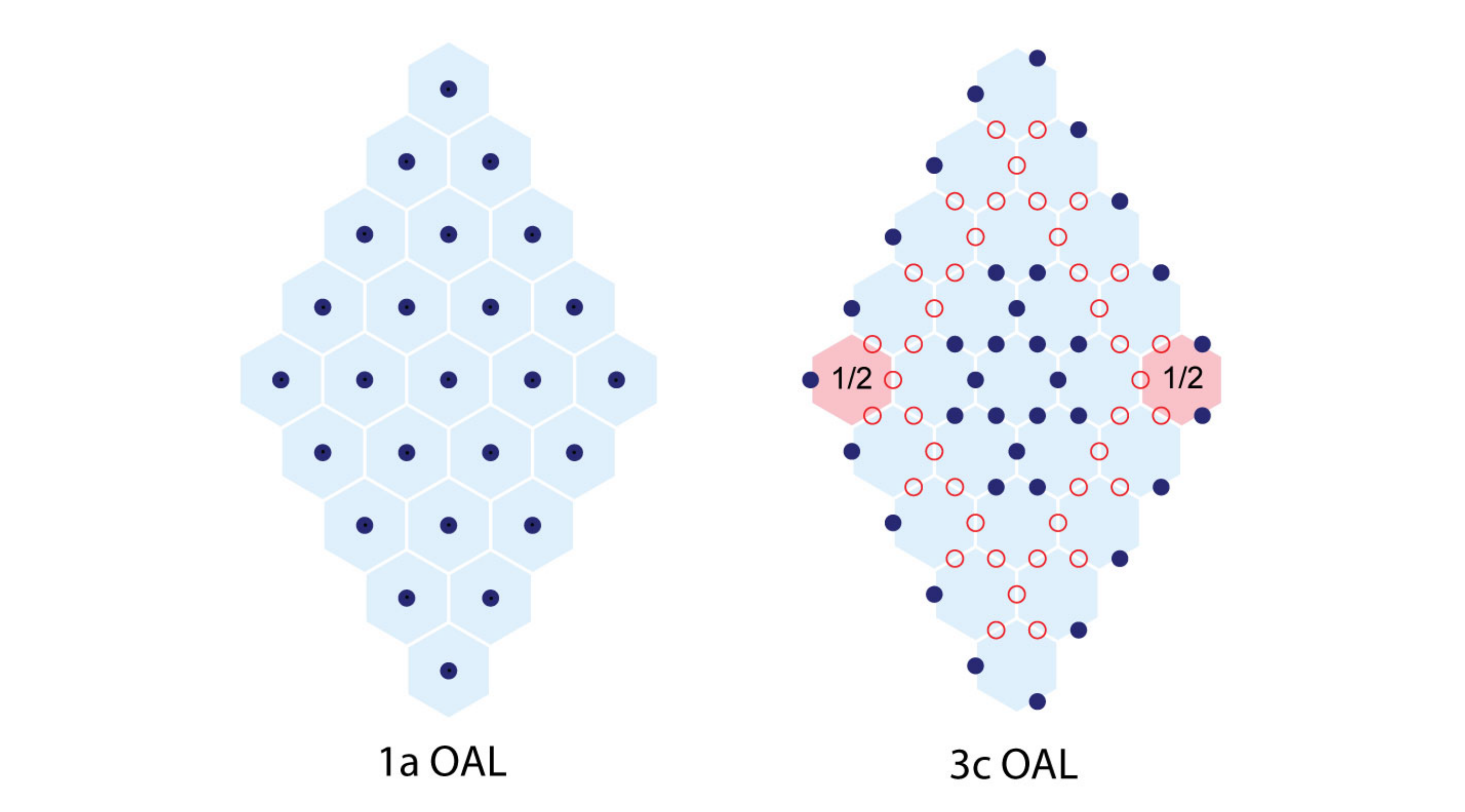}
	\renewcommand{\thefigure}{S\arabic{figure}}
\caption{Wannier centers in the trivial (left) and topological (right) phases protected by $C_6$ symmetry. In the trivial phase, there are three Wannier centers at the center of each unit cell. There is no filling anomaly. In the topological phase, the Wannier centers localize at the edges of the unit cells. Red hollow circles are Wannier centers that are shared by unit cells at the boundary of the crystal. This arrangement is such that each of the two unit cells at 120$^\circ$ corners has an odd number of shared Wannier centers with neighboring unit cells (all other unit cells have an even number of shared Wannier centers).  This distribution of Wannier centers results in a fractional local density of states at these corners (indicated by the red shaded region) above and below the energy gap, which has to be compensated by corner-localized zero-energy modes.}
\label{si:fig:WannierCenters}
\end{figure}

\begin{figure}[t]
\includegraphics[width=\columnwidth]{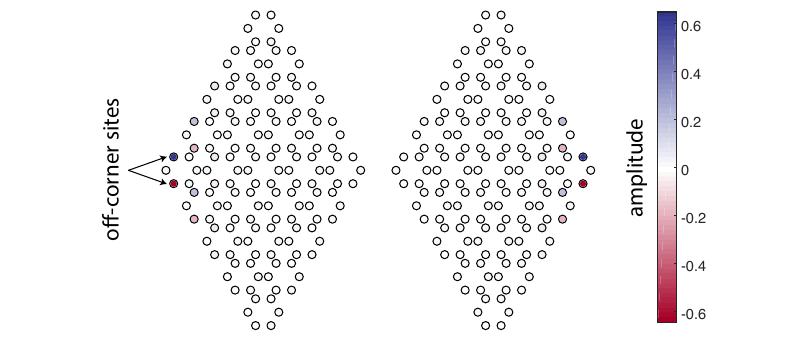}
	\renewcommand{\thefigure}{S\arabic{figure}}
\caption{Real part of the two topological zero-energy corner states in the topological phase protected by $C_6$ symmetry (imaginary part is zero. There is one state per corner, each associated with a fractional filling anomaly of $1/2$ (see Fig.~\ref{si:fig:WannierCenters}, right). Note that the amplitude of the eigenstate at the two ``off-corner" sites have opposite phases.}
\label{si:fig:TopologicalStates}
\end{figure}

As detailed in the main text, when the lattice is in the topological phase, $\chi^{(6)}_\mathcal{T}=(2,0)$, it has three Wannier centers at the edges of the unit cells (Fig.~\ref{si:fig:WannierCenters}, right). Note that it is impossible to deform the Wannier centers away from that position in a way that preserves $C_6$ symmetry. This obstruction to symmetry-preserving deformations is a real-space manifestation of the symmetry protection of the phase. For this reason, this phase is said to be in an `obstructed' atomic limit~\cite{bradlyn2017}. It was recently shown that obstructed atomic limits with no dipole moments (such as this lattice), can present a `filling anomaly' when corners are introduced. In insulators, a corner-induced filling anomaly results in the fractionalization of corner charge~\cite{benalcazar2019fillinganomaly}. In our system, the filling anomaly manifests a fractional local density of states at $120^\circ$ corners (Fig.~\ref{si:fig:WannierCenters}). In the additional presence of chiral symmetry, the filling anomaly is compensated by the existence of zero-energy corner states. These states, shown in Fig.~\ref{si:fig:TopologicalStates}, are the states that need to be excited at the beginning of the pumping cycle. These states are robust. Their energies are protected at a value of zero by chiral symmetry; their degeneracy is further protected by $C_6$ symmetry. Breaking either of these symmetries softly (i.e., without causing a bulk phase transition) will generally preserve these corner states. Indeed, adding terms that break chiral symmetry (e.g., by including next nearest neighbor couplings) will not compromise the pump, as the Wannier centers will remain fixed by $C_6$ symmetry to the edges of the unit cells. The pumping procedure described in the main text may consequently not be symmetric around $\theta=0$ or $E=0$, but the in-gap corner states will necessarily cross the gap, and therefore traverse the structure from one corner to the opposite one. 

\section{Full wave simulations}
\label{si:section:sim}
We run simulations based on the beam propagation method (BPM) to determine the experimental parameters. Here we show the result of the final simulation, with parameters used in the experimental realization. These are a full-wave simulation of the evolution of a beam $\psi(\textbf{r},z)$ in the propagation direction, $z$, using the paraxial equation Eq.~\eqref{si:eq:propagation}. 
The waveguide in the simulation is modeled as having a Gaussian profile for the variation in the waveguide refractive index:
$\Delta n(x,y) = (3.00\times10^{-3}\pm \delta n_{0}) \exp(-x^{2}/\sigma_{x}^{2}-y^{2}/\sigma_{y}^{2})$, with $\sigma_{x}=3.5$~$\mu$m and $\sigma_{y}=5.35$~$\mu$m.
Using these simulations, we optimized the parameters of modulation of the waveguide's separations and refractive indices to maximize the adiabaticity and the efficiency of the pump. For that purpose, we fixed the sample length to $z=15$ cm, which is the maximum length attainable in our experimental setup. 
The modulation pattern we obtained makes the relation between the adiabatic parameter, $\theta$, and the direction of propagation of light, $z$, piecewise linear,
\begin{align}
\theta(z) =
\begin{cases}
-\pi + \frac{\theta_{c}}{z_{c}}\,z & \;\; 0<z<z_{c}\\
(\theta_{c}-\pi) + \frac{2(\pi-\theta_{c})}{z_{L}-2z_{c}}\,(z-z_{c}) & \;\; z_{c}<z<z_{L}-z_{c}.\\
(\pi-\theta_{c}) + \frac{\theta_{c}}{z_{c}}\,(z-z_{L}+z_{c}) & \;\; z_{L}-z_{c}<z
\end{cases}
\end{align}
Additionally, the amplitude of modulation of the waveguides' separations, $A$, varies according to
\begin{align}
A =
\begin{cases}
 1.6~\mu\text{m}      & \;\; |\theta| < 0.5\pi\\
 2.3~\mu\text{m} & \;\; |\theta| > 0.5\pi
\end{cases}
\label{si:eq:modulation}
\end{align}
where $z_{L}=15$~cm is the total sample length, $\theta_{c}=0.32\pi$, $z_{c}=5.6$~cm and $s(\theta) =L/3-A\cos(\theta)$.

\begin{figure}[t]
\includegraphics[width=\columnwidth]{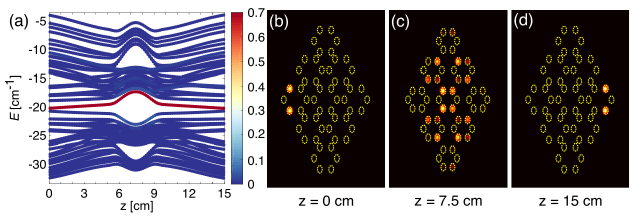}%
\caption{Simulation of beam propagation during the second-order pumping process. (a) Spectrum during one pumping cycle. The color map indicates the amplitude of the projection of the beam onto each of the instantaneous eigenstates of the system [same as Fig.~3(a) in the Main Text]. (b) Beam profile at the input facet, $z=0$, which occupies the left topological corner state. (c) Beam at half cycle, occupying the lowest bulk state above the gap due to state-level adiabaticity. (d) Beam profile at output facet, $z=15$ cm, which occupies the right topological corner state.}
\label{si:fig:simulation}
\end{figure}

Figure~\ref{si:fig:simulation} shows the simulation of a pump cycle for this modulation. Figure~\ref{si:fig:simulation}(a) plots the instantaneous energies at each depth in the array. The color indicates the projection of the beam into the instantaneous eigenstates of the system in the $(x,y)$ plane. Figures~\ref{si:fig:simulation}(b)-\ref{si:fig:simulation}(d) show the intensities of the wave function at the cross sections of the system at the beginning, middle, and end of the cycle, respectively. The initial wave function occupies the left topological corner state of the second-order topological phase [Fig.~\ref{si:fig:simulation}(b)]. As the wave function adiabatically propagates, it delocalizes into the bulk. Such delocalization is maximal in the middle of the cycle [Fig.~\ref{si:fig:simulation}(c)]. In the second half of the cycle, the beam increasingly localizes on the right corner. At the output facet, the wave function largely occupies the right topological corner state [Fig.~\ref{si:fig:simulation}(d)].

\section{Waveguide fabrication}
We fabricated the waveguides using femtosecond direct laser writing technique.
Using the optimized waveguide parameters found by BPM simulation, we wrote the waveguides using a 800~nm titanium:sapphire laser and amplifier system (Coherent:RegA 9000 with pulse duration 270~fs, repetition rate 250~kHz, and pulse energy 820~nJ) in borosilicate glass (Corning Eagle XG borosilicate glass) with refractive index $n_{0}=1.473$ at $\lambda = 1550$~nm.
The shape and size of the focal volume were controlled by first sending the laser writing beam through a beam-shaping cylindrical telescope and then focusing it inside the glass chip using a $\times$50, aberration-corrected microscope objective (NA = 0.55).
The waveguides were fabricated by translating the glass chip through the focal volume of the laser beam using a high-precision three-axis Aerotech motion stage (model ABL20020). 
The refractive index modulation of the sample was achieved by speed variation of the laser writing beam.

\bibliography{references}
\end{document}